\documentclass[aip,graphicx]{revtex4-1}

\usepackage{graphicx}
\usepackage{dcolumn}
\usepackage{bm}

\usepackage[utf8]{inputenc}
\usepackage[T1]{fontenc}
\usepackage{mathptmx}
\usepackage{etoolbox}
\usepackage{float}
\usepackage{amsmath}
\usepackage{xcolor}
\usepackage{siunitx}

\draft 

\begin{document}

\title{Optically Driven Janus Micro Engine with Full Orbital Motion Control} 

\author{David Bronte Ciriza}
\email[]{brontecir@ipcf.cnr.it}
\affiliation{CNR-IPCF, Istituto per i Processi Chimico-Fisici, I-98158, Messina, Italy}

\author{Agnese Callegari}
\email[]{agnese.callegari@physics.gu.se}
\affiliation{Department of Physics, University of Gothenburg, SE-41296, Gothenburg, Sweden}

\author{Maria Grazia Donato}
\affiliation{CNR-IPCF, Istituto per i Processi Chimico-Fisici, I-98158, Messina, Italy}

\author{Berk Çiçek}
\affiliation{Department of Mechanical Engineering, Bilkent University, TR-06800, Ankara, Turkey}

\author{Alessandro Magazzù}
\affiliation{CNR-IPCF, Istituto per i Processi Chimico-Fisici, I-98158, Messina, Italy}

\author{Iryna Kasianiuk}
\affiliation{Department of Mechanical Engineering, Bilkent University, TR-06800, Ankara, Turkey}
\affiliation{UNAM - National Nanotechnology Research Center and Institute of Materials Science \& Nanotechnology, Bilkent University, 06800 Ankara, Turkey}

\author{Denis Kasianiuk}
\affiliation{Department of Mechanical Engineering, Bilkent University, TR-06800, Ankara, Turkey}
\affiliation{UNAM - National Nanotechnology Research Center and Institute of Materials Science \& Nanotechnology, Bilkent University, 06800 Ankara, Turkey}

\author{Falko Schmidt}
\affiliation{Nanophotonic Systems Laboratory, Department of Mechanical and Process Engineering, ETH Zurich, CH-8092, Zurich, Switzerland}

\author{Antonino Foti}
\affiliation{CNR-IPCF, Istituto per i Processi Chimico-Fisici, I-98158, Messina, Italy}

\author{Pietro G. Gucciardi}
\affiliation{CNR-IPCF, Istituto per i Processi Chimico-Fisici, I-98158, Messina, Italy}

\author{Giovanni Volpe}
\affiliation{Department of Physics, University of Gothenburg, SE-41296, Gothenburg, Sweden}

\author{Maurizio Lanza}
\affiliation{CNR-IPCF, Istituto per i Processi Chimico-Fisici, I-98158, Messina, Italy}

\author{Luca Biancofiore}
\email[]{luca@bilkent.edu.tr}
\affiliation{Department of Mechanical Engineering, Bilkent University, TR-06800, Ankara, Turkey}
\affiliation{UNAM - National Nanotechnology Research Center and Institute of Materials Science \& Nanotechnology, Bilkent University, 06800 Ankara, Turkey}

\author{Onofrio M. Maragò}
\affiliation{CNR-IPCF, Istituto per i Processi Chimico-Fisici, I-98158, Messina, Italy}

\date{\today}

\begin{abstract}
Microengines have shown promise for a variety of applications in nanotechnology, microfluidics and nanomedicine, including targeted drug delivery, microscale pumping, and environmental remediation. However, achieving precise control over their dynamics remains a significant challenge. In this study, we introduce a microengine that exploits both optical and thermal effects to achieve a high degree of controllability. We find that in the presence of a strongly focused light beam, a gold-silica Janus particle becomes confined at the stationary point where the optical and thermal forces balance. By using circularly polarized light, we can transfer angular momentum to the particle breaking the symmetry between the two forces and resulting in a tangential force that drives directed orbital motion. We can simultaneously control the velocity and direction of rotation of the particle changing the ellipticity of the incoming light beam, while tuning the radius of the orbit with laser power. Our experimental results are validated using a geometrical optics phenomenological model that considers the optical force, the absorption of optical power, and the resulting heating of the particle. The demonstrated enhanced flexibility in the control of microengines opens up new possibilities for their utilization in a wide range of applications, encompassing microscale transport, sensing, and actuation.
\end{abstract}

\maketitle

\section{Introduction}

Microengines have steadily gained popularity and become prevalent as effective tools for controlling processes on small scales \cite{Fusi2023}. Their ability to convert energy into active motion makes them essential for nanotechnology applications such as generating precise fluid flows in microfluidic chips \cite{Lin2012,Butaite2019,schmidt2018microscopic}, delivering drugs more efficiently in nanomedicine \cite{wang2012cargo,guix2014nano,pellicciotta2023light}, or for environmental remediation \cite{soler2013self,Safdar2017}. Janus particles \cite{walther2008janus}, characterized by two distinct hemispheres with different physical properties, are the most widely used model system for microengines. Their inherently asymmetric design allows them to self-propel under various conditions. For instance, dielectric Janus particles can be designed with a metallic cap that generates a local, asymmetric heat profile under light exposure, resulting in its directed motion \cite{merkt2006capped,moyses2016trochoidal,nedev2015optically,ilic2016exploiting,jiang2010active}. While microengines are able to overcome random thermal fluctuations and exhibit directed motion, the lack of control over their dynamics is a significant limitation for their broader application.

Light is one of the most efficient approaches to induce and control the motion of microengines \cite{palagi2019light,buttinoni2022active,Tkachenko2023}. Although non-optical electric \cite{fan2012electronic} and magnetic fields \cite{gao2012cargo} are also promising alternatives, light has distinct advantages such as high energy density, precise control over its position and time, and the ability to effectively transfer both linear and angular momentum \cite{jones2015optical}. Specifically, a highly focused laser beam can confine particles around the focal point through the exchange of momentum between light and particles, a technique known as optical tweezers \cite{volpe2023roadmap}. Once confined, by transferring momentum to the particle, there are two main strategies for turning the trapped particle into a rotating microengine. Firstly, spin \cite{arita2013laser,friese1996optical,friese1998optical} and/or orbital \cite{simpson1997mechanical,friese1996optical,pesce2015step} angular momentum can be transferred to the particle, generating a polarization or phase-dependent torque that drives orbital rotations. The direction of rotation can be controlled by adjusting the beam's polarization or phase gradients. Secondly, for asymmetric particles the scattering generates an optical torque \cite{galajda2001complex,neves2010rotational,jones2015optical,magazzu2022investigation} where the direction of rotation is fixed by the scattering pattern (windmill effect) and determined by the particle's shape. This effect has also been observed for metal-dielectric Janus particles \cite{merkt2006capped,zong2015optically}, highlighting the relevance of both light scattering and thermal effects \cite{merkt2006capped}.

Indeed, for light absorbing particles, not only momentum transfer but also energy absorption and consequent heating plays a key role in their dynamics, giving rise to more complex behaviours \cite{franzl2021fully, maggi2015micromotors, ilic2016exploiting, nedev2015optically,merkt2006capped,girot2016motion,schmidt2018microscopic,moyses2016trochoidal,liu2016self,mousavi2019clustering, svak2018transverse}. Because of the combination of optical and thermal effects, microengines can show elevator-like motion \cite{nedev2015optically}, elliptical \cite{liu2016self}, trochoidal \cite{moyses2016trochoidal} and circular orbits \cite{merkt2006capped,girot2016motion,schmidt2018microscopic, svak2018transverse}, can rotate at higher velocities \cite{maggi2015micromotors}, and present reconfigurable assemblies of multiple particles \cite{paul2022optothermal}. This shows that the integration of optical and thermal effects can induce a diverse range of dynamic behaviours. However, the controllability over these dynamic behaviours is very limited. For instance, unless the beam is continuously repositioned \cite{franzl2021fully}, the direction of rotation is either fixed by a previously designed particle's geometry \cite{maggi2015micromotors} or is erratic and influenced by random thermal fluctuations \cite{girot2016motion,schmidt2018microscopic,merkt2006capped,liu2016self}. Thus, a more sophisticated scheme is required to simultaneously manipulate the direction of rotation and the angular velocity in order to enhance the control of microengines.

In this study, we combine the precise control obtainable via optical forces with the strong driving forces of thermal effects to realize a microengine that allows simultaneous control of its speed, radius, and direction of rotation using a single beam of light. Specifically, we investigate a gold-silica Janus particle trapped by a linearly polarized Gaussian beam at a distance from the beam's center where the opposing optical and thermal forces balance. By employing circularly polarized light, the transfer of the light's spin angular momentum to the particle induces a tilt in the particle orientation. This tilt breaks the symmetry between the optical and thermal forces acting on the particle, leading to simultaneous rotations around the beam's axis and around the particle's axis (the particle rotates with the gold side always pointing inwards). We control the particle’s direction of rotation and angular velocity by tuning the beam’s ellipticity, showing that transitions between rotational and stationary states can be achieved within the same system. The experimental results are in agreement with an extended geometrical optics phenomenological model that also considers the polarization of the light beam and enables the calculation of the optical power absorbed in the particle's cap. Our findings delve into the complexities of light-matter interactions in thermally driven microengines, presenting new insights and paving the way for enhanced control and manipulation in the field of nanotechnology.

\section{Results and discussion}

In this study, we investigate a microengine driven by both optical and thermal effects and whose motion we can precisely control by adjusting the power and polarization of the incident light beam. The microengine consists of a gold-capped silica Janus particle fabricated by sputtering a 10\,nm-thick gold layer on top of a $3\,\rm{\mu m}$ diameter silica particle (Figure~1). The particle's gold facet is optically thin enough to not drastically change its optical properties and thus its trapping capabilities but thick enough to induce thermal temperature gradients under light illumination (see Methods-Numerical Model). The beam shines from below (red arrow in Figure~1(a)) and the focal spot is located at a distance $h=8\,\rm{\mu m}$ above the particle (bright spot at the top of Figure~1(a)). When the beam is circularly polarized (white spiral in Figure~1(a)), the Janus particle performs orbital rotations at almost constant speed $v$ around the beam's center. The particle's motion is recorded via digital video microscopy at 20\,fps and tracked with customized Python routines. During its motion, the particle's gold-cap always faces inwards (vector $\vec{n}$ pointing away from the cap in Figure~1(a)) and in the presence of circular polarization is misaligned ($\theta$) with the local Poynting vector ($\vec{S}$) of the laser beam, see angle $\theta$ between the $xy$-projections of $\vec{n}$ and $\vec{S}$ (yellow and green dashed lines respectively). We observe this behaviour for various distances between particle and focal spot in the range $6 \leq h \leq 10\,\rm{\mu m}$, whereas the particle can not be trapped for $h \le 6 \,\rm{\mu m}$ or does not rotate for $h \ge 10 \,\rm{\mu m}$. We find that the microengine is driven by both optical and thermal effects, and can be precisely controlled by adjusting the power and polarization of the incident light beam. Through both experimental and numerical analysis, we explore the dynamics of the microengine under varying light power and polarization conditions.

\begin{figure*}[ht]
\label{Fig:1}
\centering\includegraphics[width=1\textwidth]{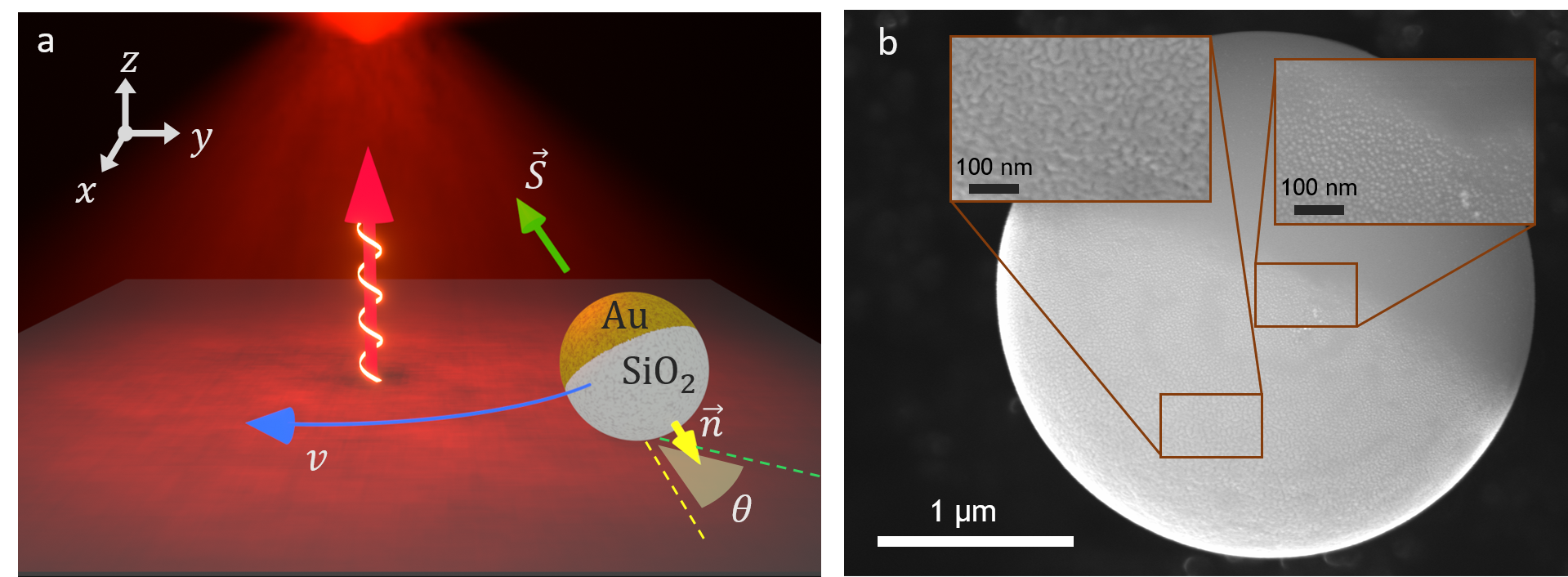}
\caption{\textbf{Orbital motion of Janus particle under circularly polarized light.}
a) Schematic of the orbital motion of a gold-capped Janus particle made of SiO$_2$ under a circularly polarized focused beam (red arrow with white spiral illustrating the 
direction and polarization of the light beam). The particle is constantly rotating at speed $v$ (blue arrow) around the center of the beam $8\,\rm{\mu m}$ below its focal point (red spot on the top). The particle's orientation is sligthly tilted at angle $\theta$, which indicates the misalignment between the $xy$-projections of the cap orientation (yellow arrow, $\vec{n}$) and the local Poynting vector (green arrow, $\vec{S}$). b) SEM images of the fabricated Janus particles. Insets show specific regions of the particle, where the left inset show deposited gold layer and the right inset the transition from the gold cap to the SiO$_2$ particle.
}
\end{figure*}

\subsection{Motion as a function of laser power}

When the light is circularly polarized, the Janus particle performs continuous circular orbits (Figure~2(a)). Upon increasing the power of the light beam we observe that both the orbital radius ($\rho$) and the confinement of the particle are increasing (Figure~2(b) and Supplementary Video~1). At low power ($P=6$\,mW) the Janus particle is mostly located in close proximity to the beam center ($\rho=2.4 \pm 0.6 \,\rm{\mu m}$) and the distribution of radial positions has a large standard deviation. At intermediate powers ($P=16$\,mW) the radius of motion increases and the radial confinement is enhanced, resulting in a narrower radial distribution. The average radial position peaks at the maximum power of our laser $P=34\,\rm{mW}$ with $\rho=7.5 \pm 0.4 \,\rm{\mu m}$, showing a well defined circular trajectory. Note that a fundamental difference between our experiment and other works on active colloids in optical potentials \cite{buttinoni2022active, fodor2021active} lays on the fact that in our case the orientation of the particle does depend on its position (gold always faces inwards).

Next, we fully characterize the dependence of the particle's motion on laser power for its change in orbital radius $\rho$, angular speed $\Omega$, and linear speed $v$. We find that $\rho$ increases non-linearly reaching the maximum radius at the maximum power ($P=34$\,mW, Figure~2(d)). Although $\Omega$ decreases slightly (between $1.6 \pm 0.2$ and $1.3 \pm 0.1 \,\rm{rad}/\rm{s}$, Figure~2(e)) the linear speed $v$ increases significantly (from $3.8 \pm 1.2$ to $9.5 \pm 0.6 \,\rm{\mu m}/\rm{s}$) with increasing laser power (Figure~2(f)). While the decrease in angular velocity with laser power is modest, the strong power dependence of the particle's radial distance is ultimately responsible for the observed increase in linear velocity.

Similarly shaped orbits such as the ones exhibited by our proposed microengine have been previously reported in the literature \cite{merkt2006capped,nedev2015optically,girot2016motion}. However, our microengine offers distinct advantages in terms of controllability. While previous systems with Janus particles in water showed sudden jumps in equilibrium position when varying laser power for circular orbits \cite{merkt2006capped} as well as for elevator-like motion \cite{nedev2015optically} our microengine exhibits a smooth dependence of the orbital radius $\rho$ with power. A similar power dependence of $\rho$ has been reported for optically heated spheres at a water-air interface (ranging between $3$ and $11 \,\rm{\mu m}$) \cite{girot2016motion}. Moreover, an advantage of our system is the presence of continuous and predictable rotations, which contrasts with the orbiting microengines reported in previous studies \cite{girot2016motion,merkt2006capped} that rotate in unpredictable directions and can change the direction of rotation randomly. Interestingly, a comparable behavior has been reported for a silica particle optically trapped in vacuum \cite{svak2018transverse} using transverse spin forces \cite{antognozzi2016direct,polimeno2018optical} instead of a combination between thermal forces and transfer of angular momentum. They found that an increase in power results in an increase in the value of $\rho$ (ranging between $0.2$ and $1.4 ,\rm{\mu m}$) and enhanced radial confinement of the particle. Both systems exhibit a better defined rotation frequency for higher powers. We can achieve this not only by increasing the power but also by increasing the ellipticity, see Figure~12 showing the analysis of the power spectral density of the particle trajectories. However, unlike our system, V. Svack et al. reported an increase of the angular velocity of the particle with power, exploiting the low-viscosity environment to achieve rotation frequencies of up to 15 kHz \cite{svak2018transverse}.

\begin{figure*}[ht]
\label{Fig:2}
\centering\includegraphics[width=1\textwidth]{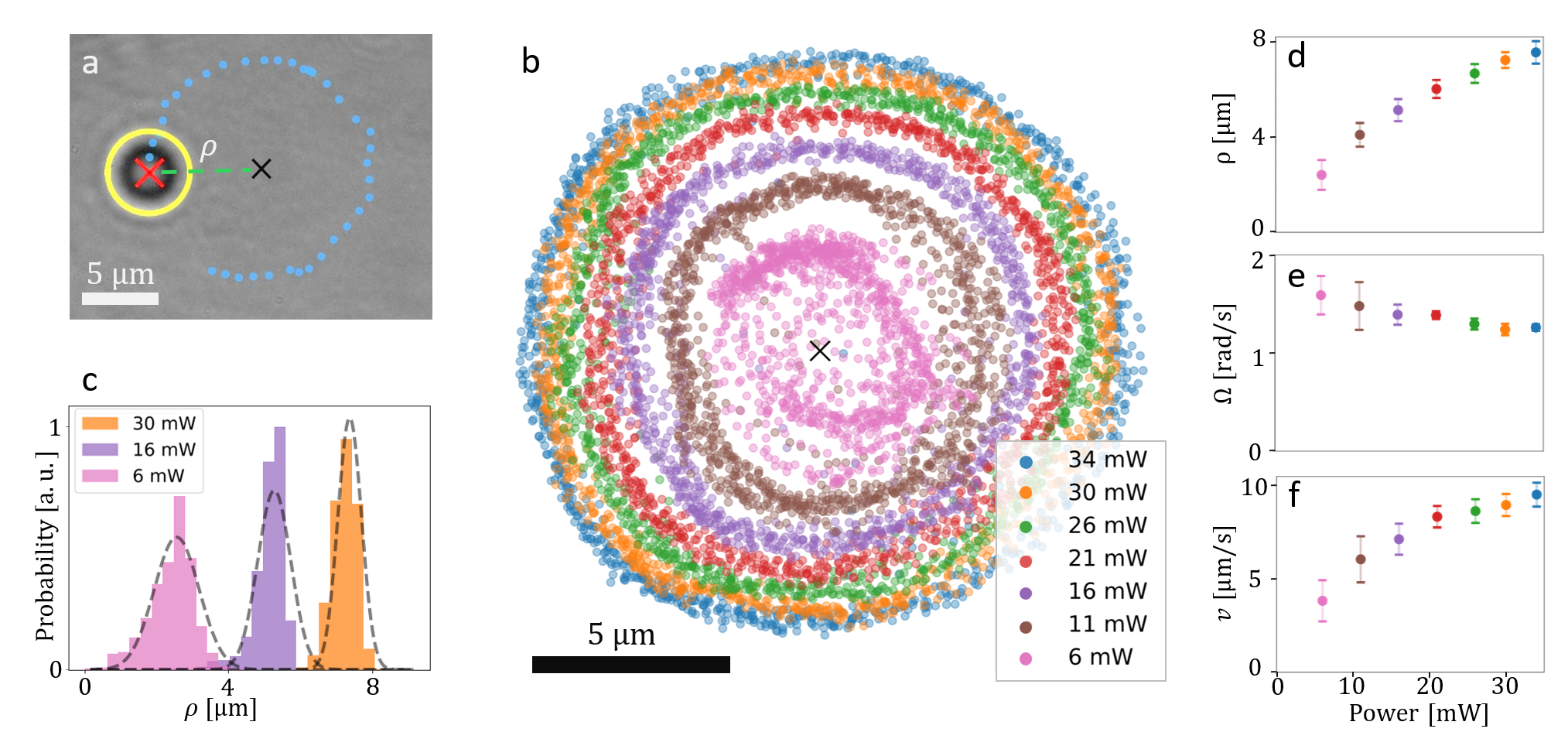}
\caption{
\textbf{Janus particle motion dependence on laser power $P$.} a) Bright-field image of a tracked Janus particle (yellow outline) with center-of-mass position (red cross) rotating around the center of the beam (marked by black cross) at a distance $\rho$ (green dashed line) with its circular trajectory (blue dotted line) during 3.5\,s. b) Recorded positions of the particle orbiting around the center of the beam (black cross) for different incident powers $P$ for 50\,s-long trajectories. c) Probability distribution of $\rho$ for three different powers ($P = 6, 16$ and $30$\,mW) and fitted with a Gaussian function (dashed grey line). From the standard deviation of the radii distribution we can quantify the particle's radial confinement. d) Average radius $\rho$, e) average angular velocity $\Omega$, and f) average linear velocity $v$ of the trajectories as a function of laser power. The error bars correspond to the standard deviation for 5 measurements of 10\,s each.
}
\end{figure*}

\subsection{Motion as a function of light polarization}
Transfer of angular momentum can induce rotation of particles around their own axis \cite{friese1998optical}. In our experiment, circularly polarized light induces a spinning rotation of the particle around its $z$-axis that breaks the symmetry between optical and thermal forces acting on it and thus induces its directional orbital motion. This motion can be stopped or reversed by changing the polarization state of the light, see Figure~3(a-c). When exposed to linearly polarized light, the particle remains confined to a specific distance $\rho$  from the center of the beam where it diffuses randomly (around the circle of radius $\rho$) due to Brownian motion, see Figure~3(b). Note that for linearly polarized light, the only acting torque is the one that orients the particle such that its gold cap ($\vec{n}$) is aligned along the local Poynting vector of the beam impinging on the particle ($\vec{S}$), see Figure~3(a-c) and Figure~11, similarly to what has been reported for a Janus particle \cite{nedev2015optically}. This alignment prevents random rotational diffusion of the particle's orientation and distinguishes our microengine from other examples in the literature where the particle rotates in random orientations \cite{merkt2006capped, girot2016motion}. When applying circularly polarized light, the direction of rotation is entirely determined by the polarization direction of the circularly polarized light, and can be reversed by switching between clockwise and anticlockwise circular polarization (Figs.~3 (a,c), and Supplementary Video~2). From the recorded video frames in Figs.~3 (a-c) we observe the gold-coated side of the Janus particle (the darkest region in transmission microscopy) facing always radially inwards to the center of the beam (yellow arrows represent the orientation vector $\vec{n}$ in Figure~1(a)). Even though the thin gold coating (10 nm) does not offer sufficient contrast to precisely quantify the exact orientation of the Janus particle, note that for circular polarization the orientation vector $\vec{n}$ is not aligned with the position vector (green dashed line in  Figure~3(a,c)) but is slightly tilted, which results in the breaking of symmetry that generates the tangential force ($\mathbf{F}_{\mathrm{tan}}$) responsible for its motion. See Figure~11 for a more detailed view of the gold-coated side and the non-coated side of the Janus particle.

Although the particle's direction of rotation is determined by the polarization of the beam, the orbit and radius of motion are independent of polarization and are solely determined by the power (as discussed in the previous section). Figure~3(d) shows the particle's positions and the direction of motion for 60\,s trajectories (each point represents 1\,s time steps). Pink and blue points represent different senses of circularly polarized light whereas orange points indicate linearly polarized light. The particle is located at a distance $\rho$ of around $7\,\rm{\mu m}$ and eventually closes a loop in approximately 6\,s. The erratic Brownian motion observed for linear polarization (orange points in the upper right corner) where the particle remains at the same location and diffusing due to Brownian motion, stands in contrast to the well-defined directional motion observed for circular polarization (blue and pink points).

\begin{figure*}[ht]
\label{Fig:3}
\centering\includegraphics[width=1\textwidth]{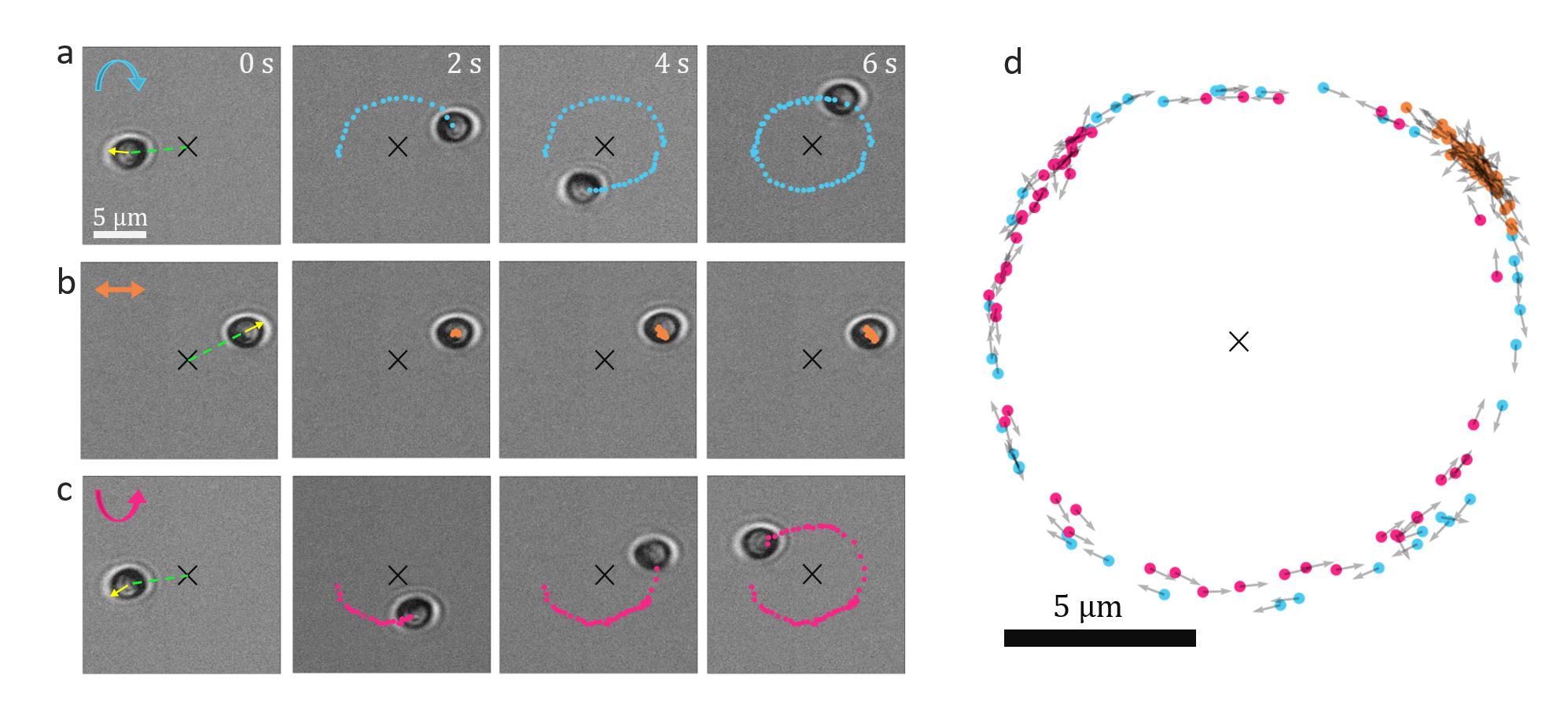}
\caption{
\textbf{Motion of a Janus particle as a function of light polarization.} a-c) The particle is shown at $t = 0, 2, 4$ and $6$\,s, the plotted points correspond to previous positions at 0.1 seconds intervals.  The black cross indicates the centre of the beam and the yellow arrows in the initial frame represent the orientation vector $\vec{n}$ as illustrated in Figure~1(a). The green dashed lines show the direction of the local Poynting vector. a) Light circularly polarized clockwise induces clockwise rotation with more than a full orbit completed after 6\,s. b) Light linearly polarized keeps the same particle at the same radius difussing with no directed motion. c) Light circularly polarized anticlockwise induces anticlockwise rotation with almost a full orbit completed after 6\,s. d) Positions (points) and direction of motion (arrows) of the particle for linearly polarized light (orange), circularly polarized light in clockwise (blue), and anticlockwise directions (pink). Positions are plotted every second for a 60\,s-long trajectory. 
}
\end{figure*}

Additional control can be gained by also adjusting the velocity and direction of rotation using elliptical polarization, as demonstrated in Figure~4. We have previously shown that changing laser power affects, both, the microengine's velocity and the radius of rotation, see Figure~2(d,f). However, adjusting the ellipticity of the light allows further velocity tuning without affecting the radius. Completely circularly polarized light yields the highest values of the angular velocity ($\Omega$), see $\phi=\pm \pi/4$ in Figure~4, where $\phi$ is the angle between the polarization plane of the linearly polarized and the fast axis of the $\lambda/4$ wave plate. Furthermore, the experimental velocities match the theoretical dependence on $\rm{sin}(2\phi)$, see Figure~4 and Supplementary Video~3. Note that the standard deviations of the angular velocities are three times larger for intermediate elliptical polarizations than for circular polarization. We attribute this to asymmetries in the beam profile profile (arising from misalignment or from the highly focused nature of the beam \cite{novotny2012principles, barnett1994orbital}) that create energy barriers that are more difficult to overcome when the tangential force $\mathbf{F}_{ \mathrm{tan}}$ is lower, resulting in a less homogeneous motion. The difference (30\%) between the experimental maximum velocity for clockwise and anticlockwise polarization is likely due to differences in the transmission of optical elements such as mirrors and dichroic beam splitters that result in slightly lower power for clockwise polarization.

Although there exist other microengines capable of producing closed orbits \cite{girot2016motion,merkt2006capped,schmidt2018microscopic}, they are unable to be stopped at a specific location within their trajectory without minimizing the power and returning to the center of the beam. Our microengine, in contrast, offers complete flexibility in terms of orbital direction, the ability to halt at any distance, and even reverse its trajectory, therefore setting a new standard in controlling microsystems that are typically dominated by random fluctuations.
The precision of control demonstrated by our proposed microengine, achieved through the ellipticity of the incoming beam, is only comparable to microengines that rely on transferring angular momentum between particles and light\cite{gao2020angular,simpson1997mechanical,friese1996optical}. However, our microengine distinguishes itself by enabling control at various distances from the beam center, rather than being limited to a single focal point.

\begin{figure*}[ht]
\label{Fig:4}
\centering\includegraphics[width=0.45\textwidth]{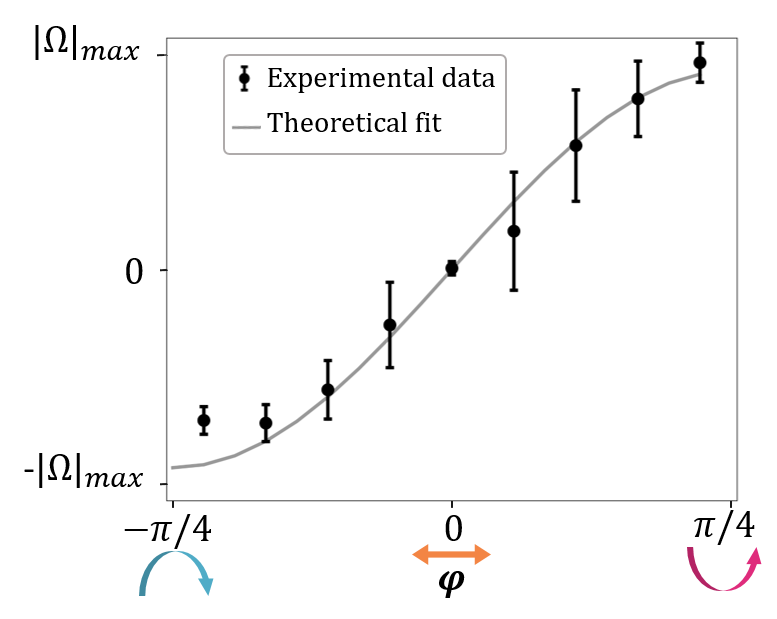}
\caption{
\textbf{Angular velocity as a function of the ellipticity of the light.} The solid line indicates the theoretical dependence on $\rm{sin}(2\phi)$. The experimental error bars correspond to the standard deviation for 5 measurements of 10 seconds each.
}
\end{figure*}

\subsection{Numerical study}

The presented micro engine is governed by a series of intricate physical phenomena. While these mechanisms possess a high level of complexity, our objective is to establish a comprehensive understanding of the driving mechanism of the system through the utilization of a simplified phenomenological model. This model incorporates three key elements. Firstly, the mechanical effects of light, which serve to attract the particle towards the center and maintain its orientation. Secondly, the light-induced heating of the particle that results in a propelling swimming force from the hot region (gold cap) to the cold region (silica part). Lastly, polarization-dependent torques that change the orientation of the particle when the light is circularly polarized. Even though our simplified phenomenological model comes with some limitations (not considering the effect of the surface in the hydrodynamics flow, assuming the gold coating to be homogeneous, not considering possible plasmonic modes…), see Methods-Numerical Model. However, it effectively captures the fundamental aspects of the experimentally studied microengine.

The model, by considering the geometrical optics approximation  \cite{callegari2015computational}, computes both the exchange of momentum between light and particle (generating optical forces \cite{jones2015optical}) and the absorption and consequent heating of the gold cap (generating thermal forces \cite{anderson1989colloid}). While the optical force draws the particle towards the center, the thermal force, caused by the difference in temperature between the gold (inner part) and silica (outer part), pushes the particle away. The combined effect of the opposing forces creates a force that cancels out at a distance $\rho$, see Figure~5(a). Furthermore, the total force has a negative slope at the point where the opposing optical and thermal forces are balanced, see the inset of Figure~5(a), leading to the formation of a stable stationary point. If the particle moves further away, it will experience a negative force that will attract the particle back towards the stable stationary position. On the contrary, if the particle approaches the center, it will experience a positive force pushing it away. For small radial displacements from the stationary position, the total radial force $F_{\rm tot}$ can be approximated as a Hookean force ($F_{\rm tot}=-k_{\rho}\,\rho$), with the stiffness $k_{\rho}$ being determined by the slope of the force in the proximity of the stationary point, see dashed red line in Figure~5(a). 

We find that our numerical analysis is consistent with our experimental results demonstrating that the orbital radius of the Janus particle increases with power, see Figure~5(b). While the forces are increasing with power, their dependence is non-linear thereby shifting the stationary position. If, both, optical and thermal forces grew linearly with the power, the stiffness would increase linearly but the stationary position wouldn't shift, as the forces would still balance at the same point, which is in contrast to experimental observations. In our model, optical forces are considered to scale linearly with the power whereas the thermal force introduces non-linearities, see Eq.~4 and Eq.~9 respectively. In our simulations, we observe a change in stationary position from $3.0 \pm 0.7 \,\rm{\mu m}$ at $10 \,\rm{mW}$ to $9.0 \pm 0.5 \,\rm{\mu m}$ at $35\,\rm{mW}$ (experiments show ranges from $2.4 \pm 0.6$ to $7.5 \pm 0.4 \,\rm{\mu m}$). Higher powers push the particle further away while increasing its radial confinement, consistently with experimental observations.

Under circularly polarized light, the transfer of spin angular momentum causes the particle to change orientation around its own $z$-axis. The force acting in the tangential direction $\mathbf{F}_{\mathrm{tan}}$ is due to the symmetry breaking between the optical and thermal forces (the optical force pulling the particle towards the center of the beam and the thermal force pushing it from the gold to the silica part). More precisely, in the presence of circularly polarized light, the orientation of the cap $\vec{n}$ is not exactly the one of the local Poynting vector $\vec{S}$, but tilted due to an additional small azimuthal rotation by the transfer of angular momentum that breaks the mirror symmetry of the configuration. This creates a steady tangential force $\mathbf{F}_{\mathrm{tan}}$ that keeps the particle rotating in its circular orbit. As we observe continuous rotations, we know that the tangential component of the thermal force should be equal to the drag force: $F_{\mathrm{tan}}=\gamma\,v$, where $\gamma$ is the viscous coefficient and $v$ is the speed of the particle (we obtain the maximum value of $F_{\mathrm{tan}}$ for maximum power and circular polarization being approximately $0.2 \mathrm{pN}$). The numerical model allows us to also determine the radial component of the thermal force (as the radius remains constant, it must have the same magnitude and opposite direction to the optical force which in this situation is approximately $1 \mathrm{pN}$). Knowing both, radial and tangential components of the thermal force, we can estimate the required tilting $\theta$ of the particle around the vertical direction (due to the torque applied by the circularly polarized light) to give the expected tangential force. We find this angle $\theta$ to be around $10$ degrees for circularly polarized light. On the other hand, when the polarization is linear (see Figure~10), the cap is aligned with the local Poynting vector such that the absence of the tangential force does not induce steady rotation but yields an equilibrium distance $\rho$ at which the particle is confined.

Our Brownian dynamics simulations (see Methods-Numerical Model) also confirm that the particle remains confined at a given radius $\rho$ that increases with power from $3.3 \pm 0.9 \,\rm{\mu m}$ at $10 \,\rm{mW}$ up to $9.1 \pm 0.6 \,\rm{\mu m}$ at $35\,\rm{mW}$, see Figure~5(c), which is consistent with both experimental and theoretical results. Additionally, the simulations verify that the radial confinement does also depend on the power, with the trajectory for higher powers being less spread than that for lower ones, see Figure~5(c). In Figure~5(d) we show simulations where different ellipticities of the incoming light such as in experiments have been considered. In particular, we plot the results for circularly polarized light in both orientations (blue and pink points) and for linearly polarized light (orange points). As for our experimental results (see Figure~3(d)), in the case of linear polarization the particle remains around the same location and only diffuses, which stands in contrast to the well-defined directional motion observed for circular polarization.

\begin{figure*}[ht]
\label{Fig:5}
\centering\includegraphics[width=1\textwidth]{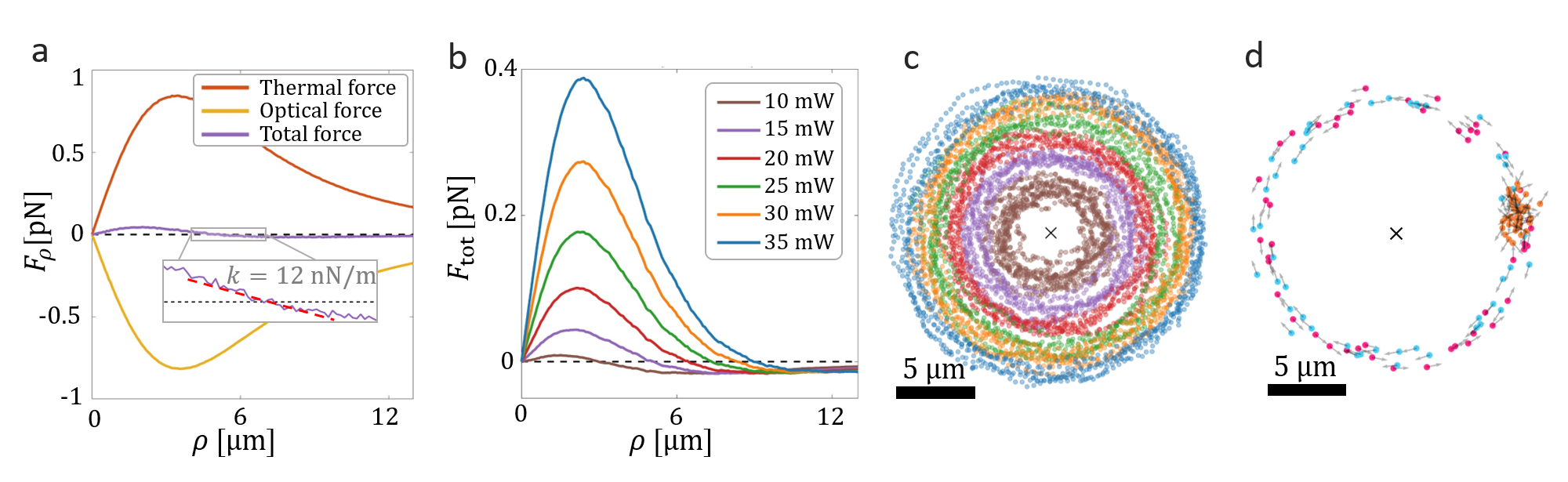}
\caption{
\textbf{Numerical study of the microengine.} a) Thermal, optical, and total force in the radial direction as a function of the radius for a power of 15 mW. The stationary point is at $ \rho=5.1\pm0.6\,\rm{\mu m}$ and the stiffness is $12\,\rm{nN/m}$. b) Total force exerted on the particle as a function of $\rho$ for different powers. Both the stationary position and the stiffness increase with the power. c) Simulation of the dynamics of the Janus particle for a 50 seconds trajectory when illuminated with different powers and d) simulation of the dynamics of the particle under anticlockwise circularly polarized light (pink), linearly polarized light (orange) and clockwise circularly polarized light (blue). The black crosses represent the center of the beam. The parameters of the plots c) and d) are identical to the ones of Figure~2 and Figure~3.
}
\end{figure*}

\section{Conclusion}

In this study, we have introduced a highly controllable microengine by combining both optical and thermal effects. We demonstrated that a $3\,\rm{\mu m}$ gold-silica Janus particle can be confined at a specific distance from the center of a highly focused beam, with the gold side facing inwards. The balance between optical forces, which pull the particle towards the high intensity region, and thermal forces, which push it away from the same region, is responsible for this confinement. Remarkably, the stationary position can be fine-tuned by adjusting the beam power. Furthermore, we showed that circularly polarized light can transfer spin angular momentum from the light to the particle, breaking the mirror-symmetry of the system and inducing a moon-like rotation (orbital motion of the particle around the beam's axis with the gold side towards the center of the beam). The speed and the orientation of this rotation can be precisely controlled by varying the ellipticity of the light. Our experimental findings have been validated by a phenomenological numerical model based on the geometrical optics approximation that matches our observations and provides further insights into the intrinsic properties of the system. Overall, the high degree of control we have achieved with this microengine opens up new possibilities in a wide range of applications, from microscale transport to sensing and actuation.

\section{Methods}

\subsection{Janus particles}
The fabrication of the Janus particles (diameter $3\,\rm{\mu m}$) made of silica ($\rm{SiO_2}$) and half coated with gold (Au) splits into three different steps. The first one consists of obtaining a crystalline monolayer of silica spheres on the glass surface. Starting from a solution of silica spheres in water, we deposit the droplet on the glass, and when the solvent evaporates we obtain a monolayer of particles on the substrate. We find the best structures when covering the substrate with a Petri-dish and keeping it at a temperature of $ 19\rm{^\circ C}$ until the sample dries. The second step consists of coating one half of the particles' surface with a 10-nm thick layer of gold. For this, we employ the thermal evaporation technique, which evaporates the metal and condenses it on the particles surface at high vacuum conditions. To improve the adhesion of gold to silica, we added a $2\,\mathrm{nm}$ layer of Titanium before adding gold. Third and last, to release the particles in solution, we immerse the substrate in water and sonicate for 5 seconds (SONICA, 1200M). SEM micrographs were collected by a Quanta 450 (FEI, Hillsboro, OR, USA) with a large-field detector (LFD) and an accelerating voltage of 20\,kV in high vacuum ($1^{-6}$\,mbar).

\subsection{Experimental setup}

To prepare the sample chamber, a small amount of Janus particles in aqueous suspension ($15-20\,\rm{\mu l}$) is drop casted on a clean microscope slide and then covered with a coverslip. The obtained chamber is sealed with nail polish to avoid evaporation during measurements. The light source for the optical tweezers is a laser diode source (Thorlabs DL8142-201) at 830 nm wavelength. After passing through a couple of anamorphic prisms and an optical isolator, the laser beam is expanded to overfill the back aperture of a high numerical aperture objective (Olympus, Uplan FLN 100X, NA=1.3), aiming at obtaining a diffraction-limited spot approximately $600\,\mathrm{nm}$ in diameter. Laser power at the objective is varied in the range between 5 and 35\,mW. A $\rm{\lambda/4}$ wave plate, placed in the beam path, is used to control the light polarization state. The relative position between the chamber and the focus of the beam is controlled using a piezoelectric stage (Mad City Labs NANO-LP200). The focal spot is located $8\,\rm{\mu m}$ above the substrate while the motion of the particle takes place directly on top of the substrate. The particle images are taken in transmission with a CCD camera and are calibrated by imaging a microscope slide ruler. Tracking of the particle dynamics follows standardized digital video microscopy techniques and has been implemented in home-made Python codes. See Figure~6 for an schematic of the experimental set up.

\subsection{Numerical model}
 
The interaction of the Janus particle with the focused Gaussian optical beam is described in the geometrical optics approximation \cite{callegari2015computational,bronte2022faster}: the beam is represented by an appropriate set of rays that, impinging on the Janus particle surface, are reflected,  transmitted, and, when hitting the gold-coated spherical cap, also partially absorbed, see Figure~7. While each ray is undergoing this infinite series of scattering events, it exchanges linear and angular momentum with the particle and therefore applies an optical force and torque. Additionally, the particle's metallic cap absorbs some of the incident light thereby increasing its temperature locally around the stationary point $\approx 5-10\ {\rm K}$. Given that the gold cap is largely continuous and gold exhibits excellent thermal conductivity, we assume the gold cap being isothermal. As the particle is immersed in solution, the temperature of the water in close proximity to the cap increases too: this asymmetry induces a temperature gradient across the particle. As fluids typically move from cold to hot regions, the particle experiences a slip flow in the opposite direction, inducing thermophoretic ($\mathbf{F}_\mathrm{thp}$) motion of the particle \cite{jiang2010active}. Moreover, the temperature increase in the volume of water close to the particle induces a volume expansion of the water ($\mathbf{F}_\mathrm{exp}$). This causes an unbalanced force towards the non-expanding volume region (i.e. the ``cold'' side of the Janus particle). In practice, the particle feels a force proportional to the increased water volume, propelling the particle towards its cold end, see Figure~8.

Our experimental observations, as depicted in Figure~3(a-c) and Figure~11, are in line with previous studies \cite{nedev2015optically} and with symmetry arguments. Based on these findings, we make the assumption that for a Janus particle with a thin gold layer, the optical torque resulting from geometrical scattering stably orients the particle in a manner where its gold cap aligns with the local Poynting vector $\vec{S}$ of the beam impinging on the particle, see Figure~1(a). While our model assumes an isothermal gold cap, the presence of small isolated grains along the cap's borders in combination with the inhomogeneous illumination induces a thermophoretic torque. This torque, similarly to the optical torque resulting from geometrical scattering, would act to align the gold cap with the local Poynting vector in the horizontal plane, thereby preserving radial symmetry. Although we did not explicitly incorporate this phenomenon into our model due to its complexity, we acknowledge its potential influence on the system's behavior. Also, the combination of particle size ($3\,\mu{\rm m}$) with the proximity to a planar boundary determines that the orientation of the particle is not very much affected by the Brownian noise, as the relaxation time of the rotational dynamics is significantly longer than the one in bulk in the order of magnitude of about $20\,\mathrm{s}$. In our dynamics simulation, hence, we consider the degrees of freedom related to the position of the particle center only, the orientation in each point defined by the local Poynting vector $\vec{S}$, see Figure~9.

From our main experimental observations we saw that elliptically polarized light induces orbital motion of the particle around the beam axis. We can simulate this by introducing a polarization dependent torque in our model. The Brownian dynamics equations for our particle, transposed already in the finite difference formalism, read as:

\begin{equation}\label{eq:langevin2Dorientation}
\left\{ 
\begin{array}{rcl}
{\Delta \rho}  & = & \displaystyle \frac{D_{\rm transl}}{k_{\rm B}T} \, F_{\rho,\rm tot}  \, {\Delta t}  + \sqrt{2 D_{\rm transl} \, \Delta t  \, }  \, W_{\rho},\\[15pt]
{\Delta s}  & = & \displaystyle \frac{D_{\rm transl}}{k_{\rm B}T} \, F_{s,\rm tot}  \, {\Delta t}  + \sqrt{2 D_{\rm transl} \, \Delta t  \, }  \, W_{s},\\[15pt]
{\Delta \psi} & = &  \displaystyle \frac{D_{\rm rot}}{k_{\rm B}T} \, T_{z,\rm pol}  \, {\Delta t},\\[15pt]
\end{array}
\right.
\end{equation}

where $\rho$ is the radial coordinate from the center of the beam and $\vec{S}$ is the coordinate in the tangential direction, oriented in the sense of positive angles (i.e., obtained from $\hat{\mathbf{\rho}}$ and the direction of the beam propagation axis $\hat{\mathbf{z}}$ via $\hat{\mathbf{s}} = \hat{\mathbf{z}} \times \hat{\mathbf{\rho}}$), and $\psi$ is the azimuthal angle describing the orientation vector of the particle in the standard lab reference frame with basis unit vectors: $\hat{\mathbf{x}}$, $\hat{\mathbf{y}}$, $\hat{\mathbf{z}}$.

The term $F_{\rho,\rm tot}$ is the total force component along the radial direction $\hat{\mathbf{\rho}}$, $F_{s,\rm tot}$ is the component along the tangential direction $\hat{\mathbf{s}}$, and $T_{z,\rm pol}$ is the torque along the beam propagation axis direction $\hat{\mathbf{z}}$ due to the amount of circular polarization of the light. The diffusion constants are $D_{\rm transl}$ and $D_{\rm rot}$, which are related to the components  $D_{||}$, $D_{\rm rot,\perp}$ of the diffusion matrix of a spherical particle\cite{mousavi2019clustering}.

The total force is calculated as:
\begin{equation}\label{eq:totalforce}
\mathbf{F}_{\rm tot} = \mathbf{F}_{\rm opt} + \mathbf{F}_{\rm thph} + \mathbf{F}_{\rm exp} + \mathbf{F}_{\rm weight} + \mathbf{F}_{\rm buoyancy} + \mathbf{F}_{\rm int}
\end{equation}
where $\mathbf{F}_{\rm opt}$ is the optical force due to the scattering of the rays on the particles,  $\mathbf{F}_{\rm thph}$ is the thermophoretic force due to the slip flow of the thin layer of fluid in the proximity of the particle surface induced by the temperature gradient along the particle diameter (direction metallic cap-uncoated end), $\mathbf{F}_{\rm exp}$ is the force due to the volume expansion of the water, caused by the temperature increase, in the region near the cap, $\mathbf{F}_{\rm weight}$ is the weight of the particle, $\mathbf{F}_{\rm buoyancy}$ is the upwards force that the fluid applies to the particle because of its mass density, and $\mathbf{F}_{\rm int}$ is the interaction force with the bottom slide, that we assume to be short range and repulsive, representing a colloidal electrostatic interaction which decays exponentially with increasing distance between the particle and bottom slide preventing sticking. As the cap is oriented in the direction of the local Poynting vector, i.e., the coated cap faces the beam focus, while the uncoated particle hemisphere faces downwards and thus the bottom slide, the vertical component of the sum of all forces except for the electrostatic interaction with the substrate is directed downwards. Therefore, we assume that the substrate must always compensate the vertical forces with the right amount of repulsion, and the particle always remain close to the substrate at a given minimal distance from it ($\approx 50\, {\rm nm}$). For this reason, we do not include an explicit equation for the particle position in the vertical direction, see Figure~10 for an schematic of the direction of the forces under different polarization conditions. Note that the presence of a surface, such as a bottom slide, can alter the hydrodynamic flows and impact the propulsion of the particle \cite{bregulla2019flow}. Although this factor may have a significant role in certain systems and the interaction between particles, we did not account for this effect in our numerical model. Instead, we deliberately developed the simplest numerical model that accurately captures the experimental observations.

The expression for the different forces are given here below. The optical force is calculated in the standard way from the scattering, summing the contribution of the force due to the single rays \cite{callegari2015computational}:
\begin{equation}\label{eq:opticalforcebeam}
\mathbf{F}_{\text{opt}}=\sum_m \mathbf{F}_{\text{ray}}^{(m)}
\end{equation}
with  
\begin{equation}\label{eq:opticalforceray}
\mathbf{F}_{\text {ray }}=\frac{n_{\mathrm{m}} P_{\mathrm{i}}}{c} \hat{\mathbf{i}}-\frac{n_{\mathrm{m}} P_{\mathrm{r}}^{(1)}}{c} \hat{\mathbf{r}}_{1}-\sum_{j=2}^{\infty} \frac{n_{\mathrm{m}} P_{\mathrm{t}}^{(j)}}{c} \hat{\mathbf{t}}_{j},
\end{equation}

The temperature increase is calculated while calculating the scattering, calculating the power absorbed by each single ray and summing it:
\begin{equation}\label{eq:temperatureincreasebeam}
P_{\rm abs}=\sum_m \mathbf{P}_{\text{cap,ray}}^{(m)}
\end{equation}
If we consider the cap isothermal, the temperature increase $\Delta T_{cap}$ is
\begin{equation}
\Delta T_{cap}=\frac{P_{\rm abs}}{(2 \pi+4) \kappa_{\mathrm{m}} R}, 
\label{eq:deltaT1}
\end{equation}
where $\kappa_{\mathrm{m}}$ is the thermal conductivity of the medium and we can define a temperature gradient across the particle given by:
\begin{equation}
\nabla T=\frac{\Delta T_{cap}}{\pi R}.
\label{eq:deltaT2}
\end{equation}
The thermophoretic velocity is expressed as $v_{\rm ph} = -D_{\rm T}\, \nabla T$ where $D_{\rm T}$ is the thermal diffusion coefficient \cite{anderson1989colloid}. From ${v}_{\rm ph}$ we obtain  ${F}_{\rm thph} = \frac{k_{\rm B}T}{D_{\rm transl}}{v}_{\rm ph}$. This force is assumed to push the particle in the $\vec{n}$ direction, from its coated cap to its uncoated end. 

The magnitude force related to the volume expansion of the water when the temperature is increased (${F}_{\rm exp}$) is modelled as follows. We estimate a linear expansion coefficient $c_{\rm L}$ for the water between the base temperature ($T$) and the increased value of the temperature ($T+\Delta T$) as:
\begin{equation}\label{eq:linearexpcoeff}
c_{\rm L} = \frac{\rho_{\rm water}(T)}{\rho_{\rm water}(T+\Delta T)} - 1,
\end{equation}
where $\rho_{\rm water}(T)$ indicates the mass density of water at  temperature $T$ and we write:
\begin{equation}\label{eq:volumeexpansionforce}
{F}_{\rm exp} = \alpha\, p_{\rm water}\, c_{\rm L}\, R^{2},
\end{equation}
where $R$ is the radius of the particle, $ p_{\rm water}$ is the hydrostatic pressure in the fluid, that we assume equal to the atmospheric pressure at sea level, and $\alpha = 0.003$ is a proportionality constant. This is a phenomenological, simplified model of the complex fluid dynamics occurring inside the fluid chamber, that are normally modelled using the Navier-Stokes equations. $\mathbf{F}_{exp}$ results from a force unbalance between the expanded water region (i.e., close to the gold cap) and the unexpanded water region (i.e., close to the silica half) and it is assumed to push the particle along the direction from its coated cap to its uncoated end. 

The polarization torque is calculated summing the contribution of each ray impinging on the particle as:
\begin{equation}\label{eq:polarizationtorquebeam}
T_{\rm pol}=\sum_m \mathbf{T}_{\text{pol,ray}}^{(m)}
\end{equation}
The contribution of each ray is modelled as proportional to the power absorbed on the first scattering event that involves the cap:
\begin{equation}\label{eq:polarizationtorqueray}
\mathbf{T}_{\text{pol,ray}}^{(m)} = \sigma \frac{ {P}_{1,\text{abs,ray}}^{(m)} }{\omega} \hat{\mathbf{r}}_{1,\text{abs,ray}}^{(m)}
\end{equation}
In the equation above, $\omega = 2\pi \nu$ is the angular frequency of the optical wavelength used for the laser beam, $\sigma$ is a parameter between -1 and 1 describing the amount of circular polarization transported by each ray (where 0 corresponds to linear polarization), ${P}_{1,\text{abs,ray}}^{(m)}$ is the power that the $m^{\rm th}$ ray deposits on the cap the first time it hits the cap, and $\hat{\mathbf{r}}_{1,\text{abs,ray}}^{(m)}$ is the direction of the $m^{\rm th}$ ray when this event happens.

\section{Data availability}
Data underlying the results presented in this paper are not publicly available at this time but may be obtained from the authors upon reasonable request.

\section{Acknowledgements}
We acknowledge financial support from the European Commission through the MSCA ITN (ETN) Project “ActiveMatter”, Project Number 812780 and the Turkish National Research Agency (TUBITAK)-CNR Bilateral project 2020-2022 “Self-assembly of complex shaped active particles in controlled optical potentials”, TUBITAK project number 119N461. D.B.C, M.G.D., A.M., A.F., P.G.G. and O.M.M acknowledge funding from the European Union (NextGeneration EU), through the MUR-PNRR project SAMOTHRACE (ECS00000022) and PNRR MUR project PE0000023-NQSTI. FS acknowledges funding from the SNSF Postdoctoral Fellowship under grant agreement 209765.

\section{Author contributions}

D.B.C. and M.G.D. performed the experiments. A.C., B.C., and D.B.C. prepared the numerical model, D.B.C. analyzed the data and wrote the initial draft. I.K. and D.K. fabricated the Janus particles. M.L. took the SEM images of the particles. L.B. and O.M.M. proposed and supervised the project. All authors discussed and commented on the results and on the manuscript text.

\section{Competing interests}
The authors declare no conflicts of interest.

\newpage

\section{Supplementary Information}

\begin{figure*}[ht]
\label{Fig:S1}
\centering\includegraphics[width=0.5\textwidth]{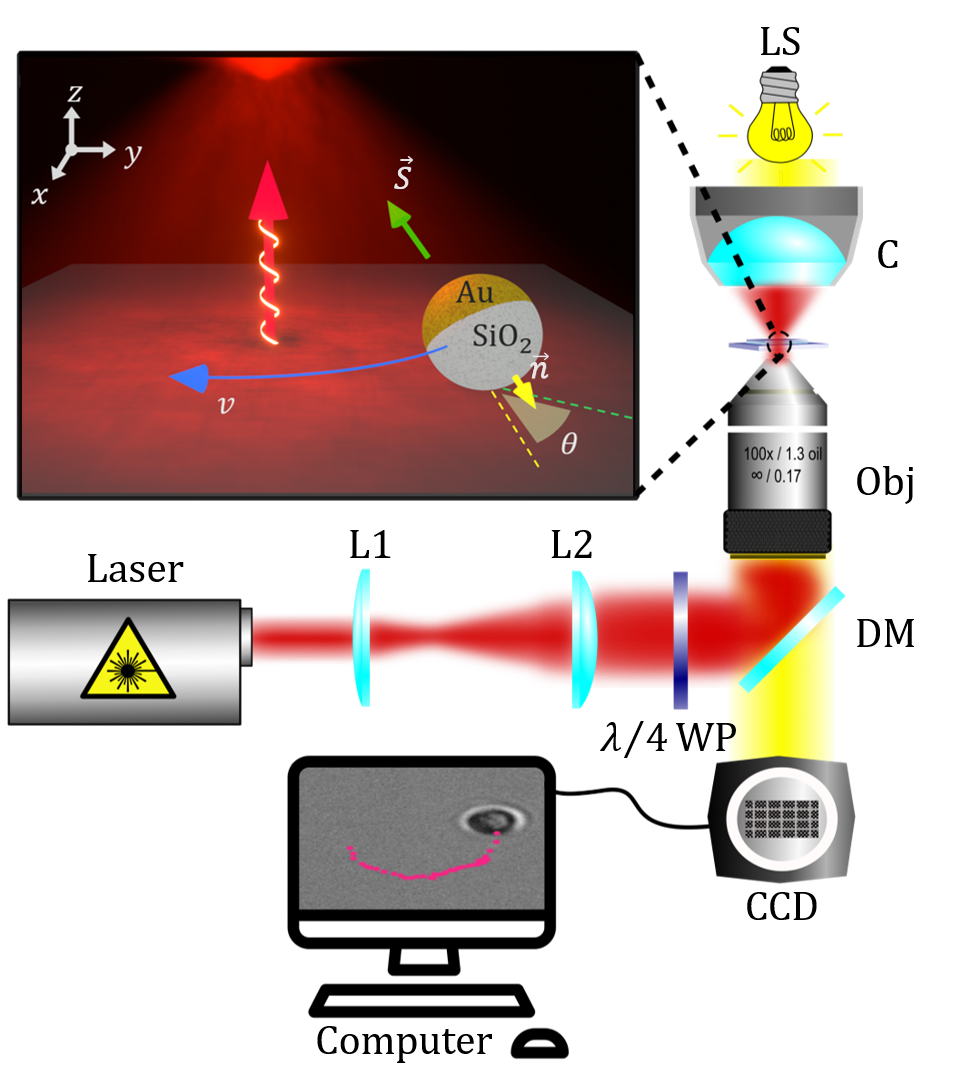}
\caption{
\textbf{Schematic of the experimental set up.} The schematic includes the laser source, the two lenses (L1 and L2) that expand the beam, the $\mathrm{\lambda/4}$ waveplate, the dicroic mirror (DM) that reflects the beam into the objective (Obj), the condenser (C), the illumination source (LS), and the computer that analyzes the images. 
}
\end{figure*}

\begin{figure*}[ht]
\label{Fig:S2}
\centering\includegraphics[width=0.5\textwidth]{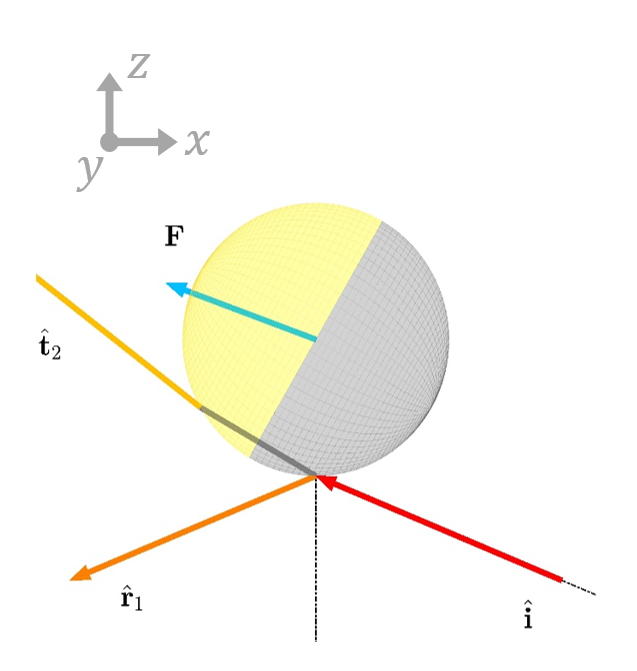}
\caption{
\textbf{Schematic of a ray impinging on a Janus particle.}
The ray $\hat{\mathbf{i}}$ reaches the particle and divides into a scattered ray $\hat{\mathbf{r}}_1$ and a transmitted ray $\hat{\mathbf{t}}_2$. The change in linear momentum results in an applied force on the particle $\mathbf{F}$
}
\end{figure*}

\begin{figure*}[ht]
\label{Fig:S3}
\centering\includegraphics[width=1\textwidth]{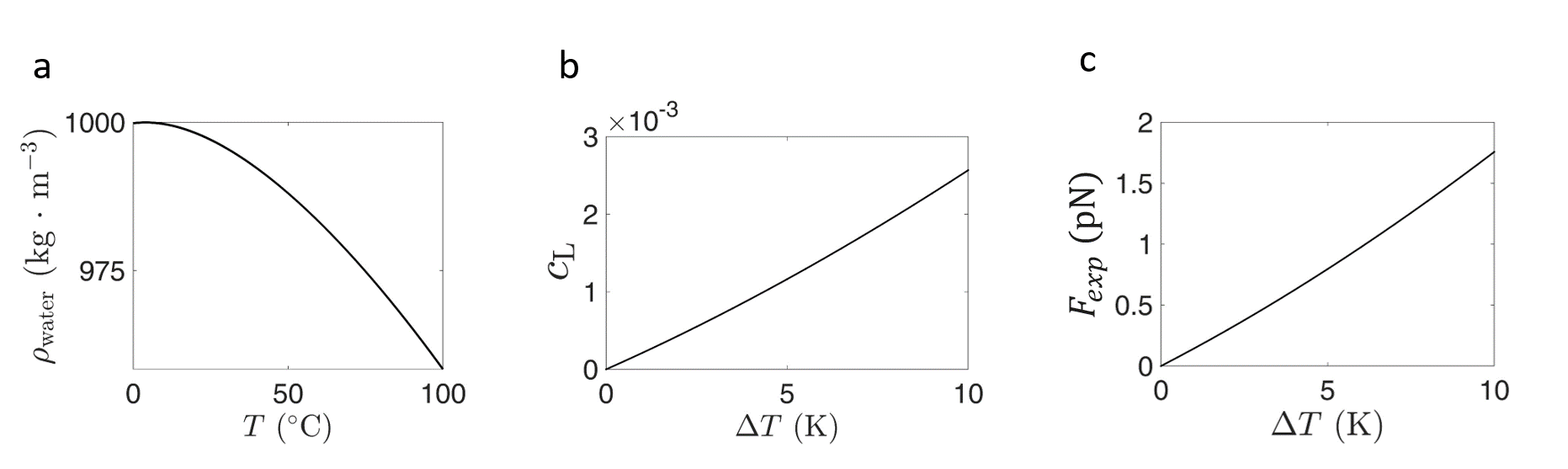}
\caption{
\textbf{Density of water ($\rho_{water}$), linear expansion coefficient of water ($c_L$) and volume expansion force ($F_{exp}$) as a function of temperature.}
}
\end{figure*}

\begin{figure*}[ht]
\label{Fig:S4}
\centering\includegraphics[width=0.5\textwidth]{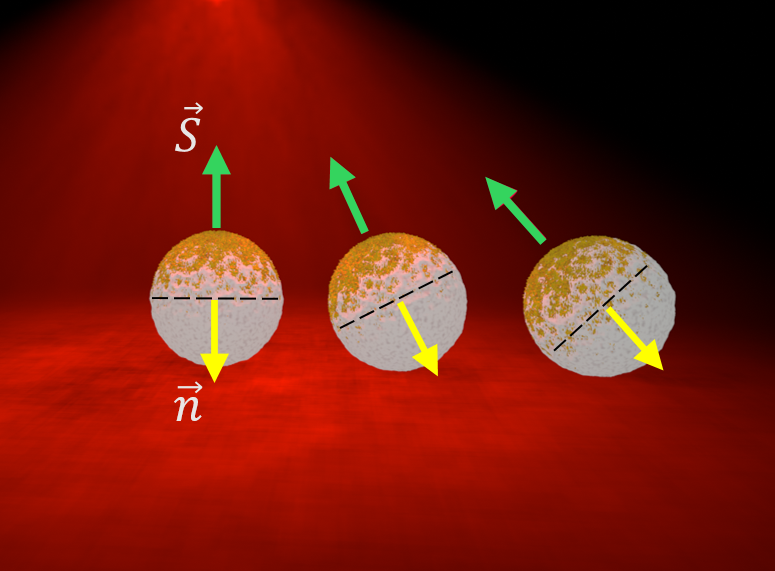}
\caption{
\textbf{Orientation of the Janus particle for different radial positions under linearly polarized light.}
The local Poynting vector of the focused beam ($S$, in green) is perpendicular to plane that contains the border between gold and silica and goes in the opposite direction to the the orientation vector ($\vec{n}$, in yellow).
}
\end{figure*}

\begin{figure*}[ht]
\label{Fig:S5}
\centering\includegraphics[width=1\textwidth]{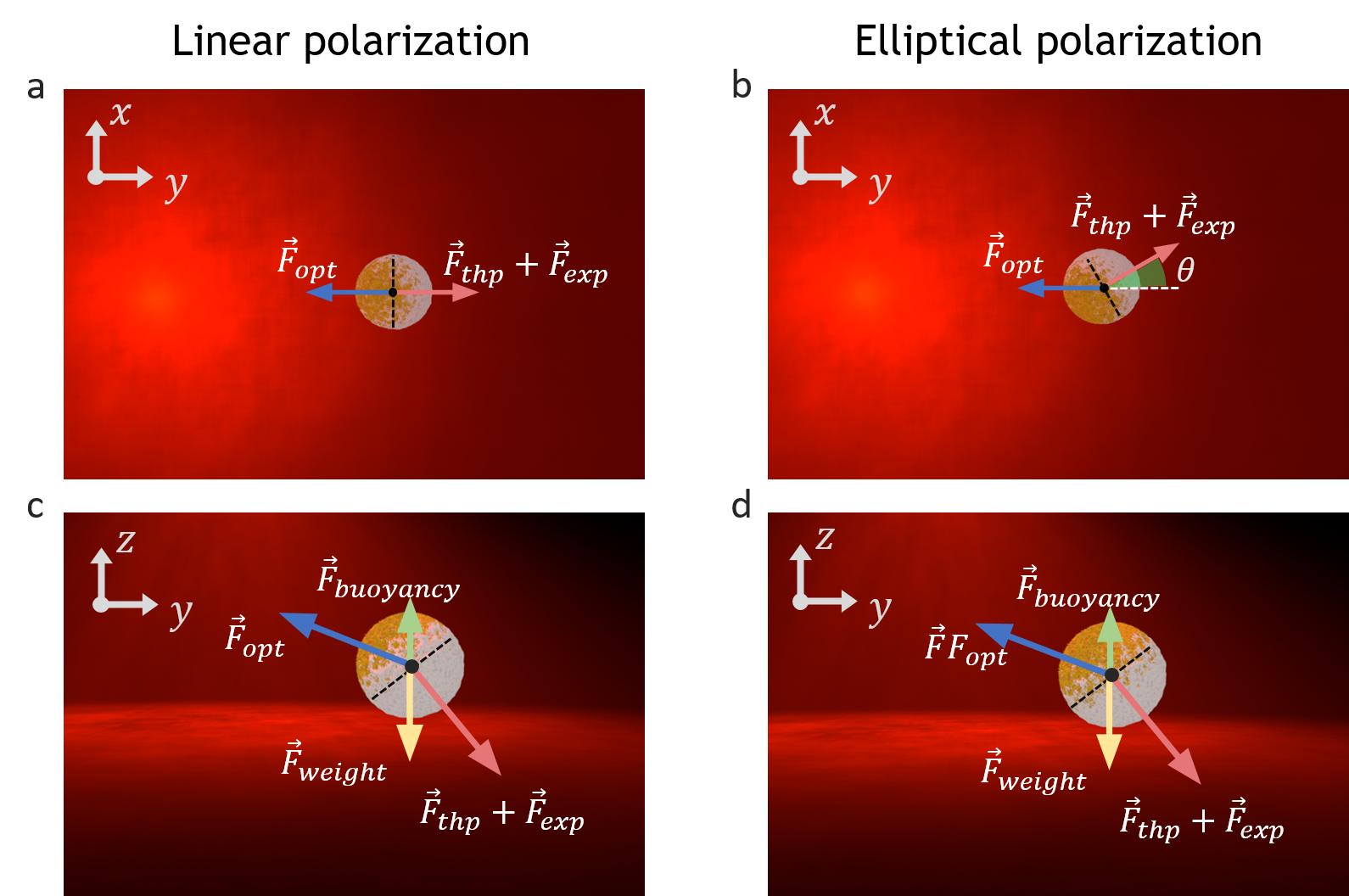}
\caption{
\textbf{Forces acting on the Janus particle under different laser polarizations.}
Forces acting on the Janus particle under linearly polarized light (a,c) and elliptically polarized light (b,d). The optical force acting on the particle can be decomposed into two components: a gradient part that directs the particle towards the center, and a scattering part in the $z$ direction, which pushes the particle away from the cover slip. In (b), $\theta$ is the angle between the direction of the optical force (towards the center of the beam) and the direction of the thermal force (from gold to silica) that gives rise to the tangential force that drives the orbital motion.
}
\end{figure*}

\begin{figure*}[ht]
\label{Fig:S6}
\centering\includegraphics[width=0.4\textwidth]{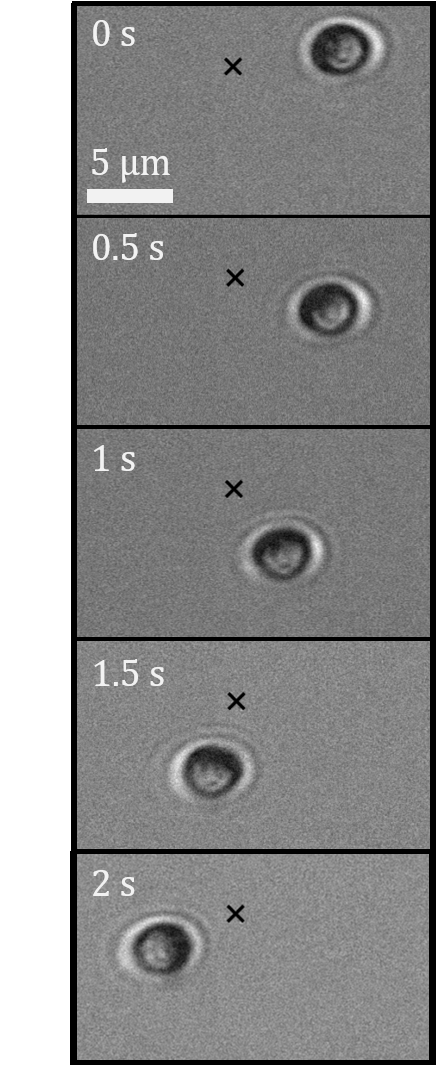}
\caption{
\textbf{Orientation of the Janus particle under circularly polarized light.}
Janus particle rotating under light circularly polarized clockwise. The particle is shown at $t = 0, 0.5, 1, 1.5$ and $2$ s. The black cross indicates the centre of the beam. The gold-coated side of the Janus particle (the darkest region in transmission microscopy) faces always radially inwards to the center of the beam while the silica particle (lightest region) faces outwards.
}
\end{figure*}

\begin{figure*}[ht]
\label{Fig:S7}
\centering\includegraphics[width=1\textwidth]{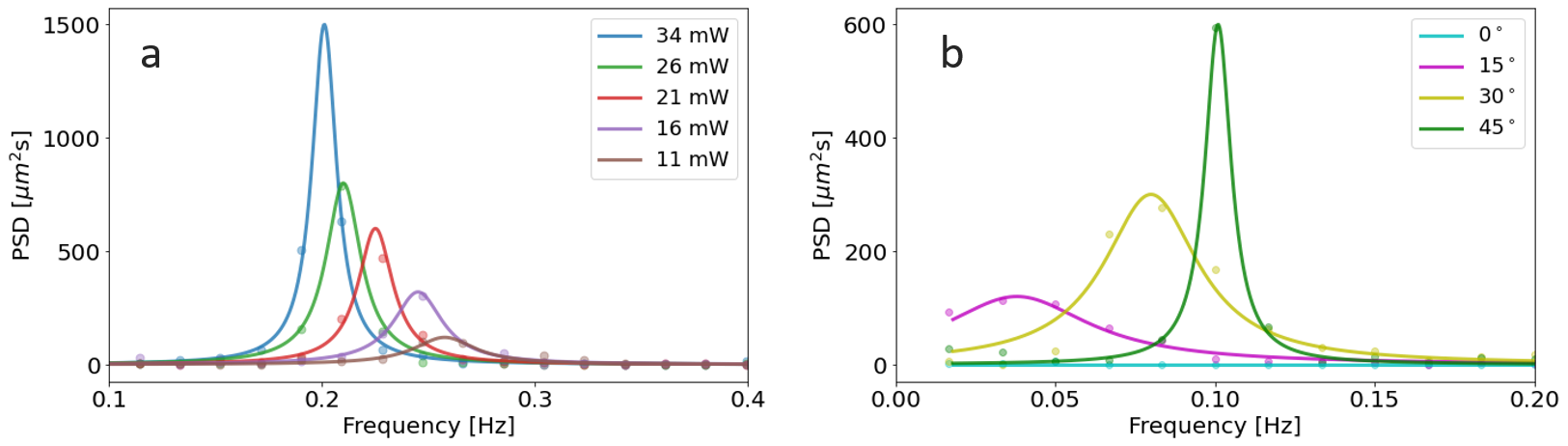}
\caption{
\textbf{Power spectral density of the trajectories of the microengine.}
(a) Light circularly polarized and with different powers. (b) Light with constant power (34 mW) and different degrees of ellipticity of the incoming light.
}
\end{figure*}

\clearpage

\bibliography{Bibliography}

\end{document}